\documentstyle[]{article}
\begin{document}

 \def\today{\number\day\enspace
      \ifcase\month\or January\or February\or March\or April\or May\or
      June\or July\or August\or September\or October\or
      November\or December\fi \enspace\number\year}
 \def\clock{\count0=\time \divide\count0 by 60
     \count1=\count0 \multiply\count1 by -60 \advance\count1 by \time
     \number\count0:\ifnum\count1<10{0\number\count1}\else\number\count1\fi}
 \def\datestamp{{\hss \today \quad\quad \clock}}
 \def\newline{\hfil\break}
\def\new{{\bf NEW: \rm }}
\def\gtsim{{_>\atop{^\sim}}}
\def\ltsim{{_<\atop{^\sim}}}
\def\et{{\it et~al. }}
\def\eg{{\it eg: }}
\def\avL{\langle \ell \rangle}
\def\avldet{\langle l_{det} \rangle}
 \def\solar{\ifmmode_{\mathord\odot}\else$_{\mathord\odot}$\fi}  
 \def\bul{\ifmmode\bullet\else$\bullet$\fi}  
 \def\avg#1{\langle #1 \rangle}
\def\xvec{{\bf x}}
\def\vvec{{\bf v}}
\def\avec{{\bf a}}
\def\mass{{\cal M}} 

\def\grad{{\nabla}}
\def\del{{\nabla}}

 \def\deg{\ifmmode^\circ\else$^\circ$\fi}
\def\arcsec{''\hskip-3pt .}
\def\arcmin{'\hskip-2pt .}
 \def\kms{\ifmmode \hbox{ \rm km s}^{-1} \else{ km s$^{-1} $}\fi} 

\def\sec{\ifmmode \hbox{\rm sec}\else{sec}\fi} 
\def\yr{\ifmmode \hbox{\rm yr}\else{yr}\fi} 
\def\myr{\ifmmode \hbox{\rm Myr}\else{Myr}\fi} 
\def\gyr{\ifmmode \hbox{\rm Gyr}\else{Gyr}\fi} 

\def\hz{\ifmmode \hbox{\rm Hz}\else{hz}\fi} 
 
\def\kpc{\ifmmode \hbox{\rm  kpc}\else{kpc}\fi} 
 \def\pc{\ifmmode \hbox{\rm  pc} \else{pc}\fi} 
\def\mpc{\ifmmode \hbox{\rm Mpc} \else{Mpc}\fi} 

\def\erg{\ifmmode \hbox{\rm erg} \else{erg}\fi} 

 \def\yrs{\ifmmode \hbox{\rm yrs}\else{yrs}\fi} 
 \def\Msun{M_\odot}
 \def\mbh{M_\bullet}
 \def\lbulge{L_{bulge}}
 \def\mbulge{M_{bulge}}
 \def\Lsun{L_\odot}
 \def\rsun{r_\odot}
 \def\angstr{\ifmmode{\rm \AA}\else\AA\fi}

\def\ho{\ifmmode H_0\else$H_0$\fi}
\def\omo{\ifmmode\Omega_0\else$\Omega_0$\fi}
\def\to{\ifmmode T_0\else$T_0$\fi}
\def\h-1{h^{-1}}
\def\rhobar{\langle\rho\rangle}
\def\dd#1#2{{d #1 \over d #2 }}  
\def\d2#1#2{{d^2#1 \over d#2^2}}
\def\rdot{\dot r}
\def\rddot{\ddot r}


\def\apj{ApJ}
\def\apjl{ApJ Letters}
\def\Apj{ApJ}
\def\aj{AJ}
\def\aa{Astron \& Astroph}
\def\annrev{Ann Rev Astron and Astroph}
\def\mn{MNRAS}
\def\pasp{Pub.~A~S~P}

\def \lap
 {\mathrel{\hbox{\raise0.3ex\hbox{$<$}\kern-0.75em\lower0.8ex\hbox{$\sim$}}}}

\def\new{{\bf NEW: \rm }}

\input psfig.tex

\title{Supermassive Black Holes Then and Now
\footnote{To appear in The Proceedings of the Second 
International LISA Symposium on Graviational Waves, 
ed. W. Folkner}}

\author{D. Richstone \\
Institute of Advanced Study, Princeton, \\
 and Dept. of Astronomy, University of Michigan}


\maketitle

\begin{abstract}
Recent surveys suggest that most or all normal galaxies 
host a massive black hole with 1/100 to 1/1000 of the visible 
mass of the spheroid of the galaxy.  Various lines of argument 
suggest that these galaxies have merged at least once in our 
past lightcone, and that the black holes have also merged.  This 
leads to a merger rate of massive black holes of about $1/\yrs$.  
\end{abstract}

\section*{Introduction}

Supermassive black holes have been a prime candidate for the probable
energy sources of quasars, the most energetic objects in the universe,
since the discovery of quasars.  Over the last decade local surveys
have suggested that quasars are present in most galaxies in the
present universe \cite{nature,kr}.  The demographics of
these objects are so fundamental to an estimate of their merger rate
that we repeat the key points below.

Except where noted, All quantities in this paper are computed for a
Friedman-Robertson-Walker Universe with $\Omega = 1 $ and $\ho = 80
\kms \mpc ^{-1}$. Distances to nearby MBHs come from many
sources, but are always rescaled to this Hubble constant. 

\section*{Statistics of Massive Black Holes in Galaxy Centers}

In about 15 cases, high resolution spectroscopy and imaging, 
coupled with detailed modelling has led to clear evidence 
for the presence of a massive dark object (MDO) in the centers 
of nearby galaxies (including our own).  The common denominator 
in all these studies is the identification of test particles (stars 
or gas clouds), which orbit the object of mass ($M$) at a distance 
$r$ at a speed $v$ given by $v^2 = \alpha \, G M /r$.  The estimate 
of $\alpha$ requires a detailed model, but often $\alpha \sim 1 $.  
In the fortunate cases of a disk of stars or gas, the analysis is 
straightforward and fairly unambiguous.  In the more complicated 
case of an anisotropic distribution of stellar orbits it is necessary  
to construct a detailed model.  The favored technique is based 
on orbit superposition and is summarized in a number of articles
\cite{maxent1,vdmm32}.  
These well-defined cases are listed in \cite{nature} and   
labelled in Figure 1. 

\begin{figure}
\vskip -0.5in
\hskip 0.3in\vbox{\psfig{figure=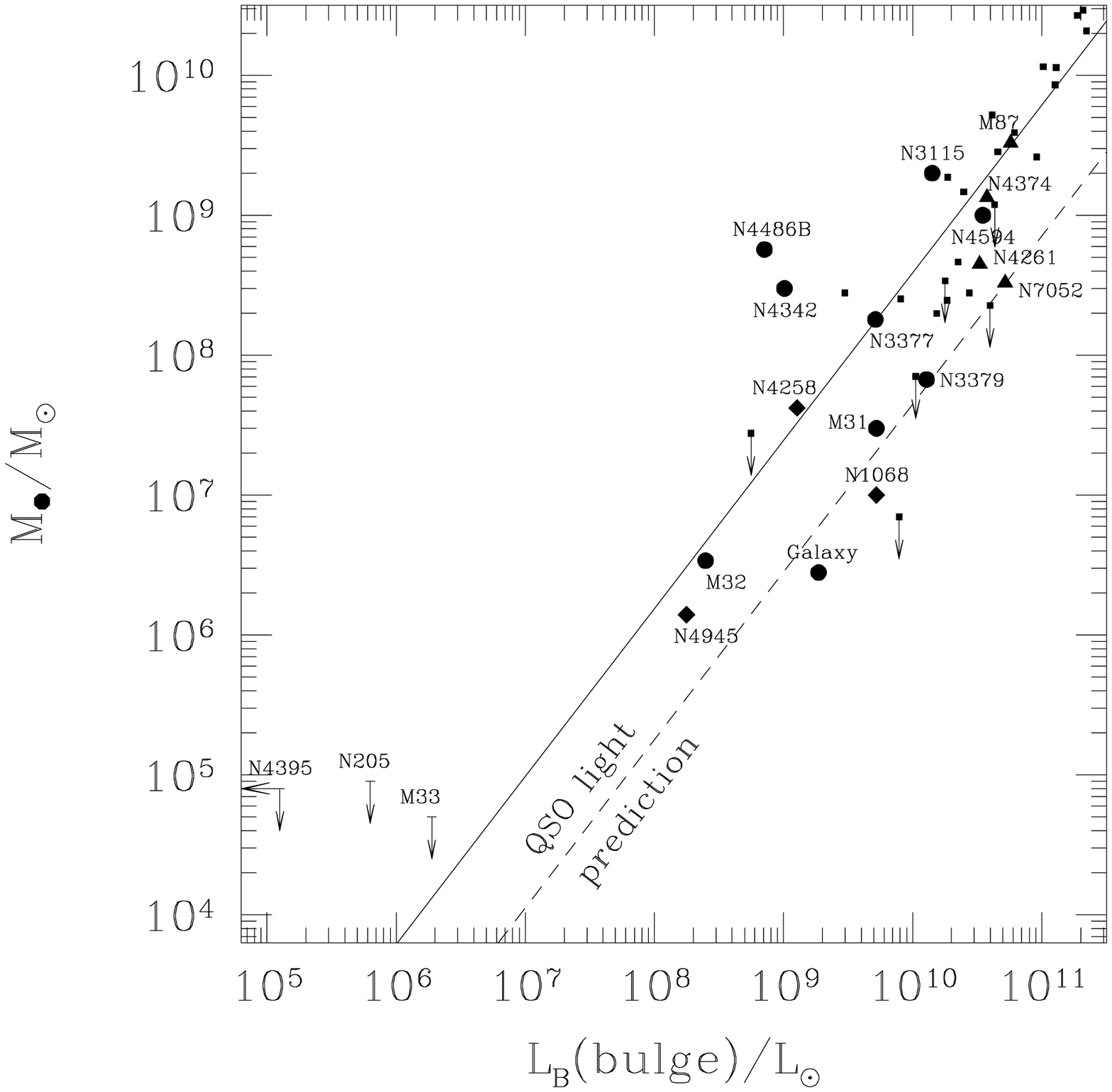,height=5.in,angle=-0.}}
\vskip -0.25in

Figure 1. ---

Mass estimates of the candidate MBHs in galaxies with dynamical
information plotted against the bulge luminosity of their host galaxy.
The labeled points are the results of painstaking observation and
detailed modelling. 
The symbols indicate the how $\mbh$ was derived: 
kinematics of gas --- triangles; dynamics of stars --- filled circles;
masers --- diamonds; or two-integral modelling using ground-based
stellar kinematics --- small squares.  Arrows indicate upper limits on
$\mbh$.  The solid line is a model with $\mbh = 0.005M_{bulge}$ 
and $ M_{bulge} = 5 \times 10^9 \Msun (L_{bulge}/10^9 \Lsun)^{1.2}$.  
The distribution of ~$\mbh$ is roughly Gaussian in
$\log (\mbh/M_{\rm bulge})$ with mean $-2.27$ ($\mbh/M_{bulge} =
0.005$) and standard deviation 0.5. 
The dashed line is the quasar light prediction of eqn 3 apportioned
according to the bulge mass: $ \mbh = 2 \times 10^7 ( L_{bulge}/5
\times 10^9 \Lsun)^{1.2}$.

The small offset from the observed
black-hole/bulge-mass relation indicates that the present integrated
density in MBHs is broadly consistent with the integrated
luminosity produced by AGNs over the life of the Universe.  This
offset may reflect a radiative efficiency of average quasar accretion
less than 0.10.  This figure is reproduced from reference 1.  

\end{figure}

In three cases (NGC 4258, the Galaxy and M32), it is possible to
reject many alternatives to a black hole (such as clusters of neutron
stars or black holes) for the observed MDO.  The basic argument
\cite{goodlee,maoz} against aggregate models is that the requirement
that the evaporation time be less than the probable system age (a
Hubble time) sets an upper limit on the mass of the constituent
objects that in these cases is near $0.1 \Msun$.  Brown dwarfs or
planets (or white dwarfs) of this mass or less would rapidly merge. 
There are no known stellar remnants of any sort of this mass.  The
MDO's might be clusters of low mass black holes or uninteracting
elementary particles (of an unknown variety), but the formation of the
former and the collapse of clusters of either to a dense state would
both require major new theories.  Based on these three objects, for
the rest of this paper we assume that all MDO's are massive black
holes (bh).

In addition to these very carefully studied cases, we
\cite{demogpaper} have combined HST images with ground--based 
spectra to analyze another 20 objects using two--integral 
distribution function based methods (this is inherently riskier
than the orbit superposition methods used for the better data).   
Combining these analyses suggest that every normal galaxy has a 
massive black dark object at the present epoch, and that the black 
hole mass is proportional to the bulge mass of the galaxy (the visible
mass of the entire galaxy if the galaxy is an elliptical).  The 
relation between bh mass and the bulge luminosity is 
 $ \mbh = 2 \times 10^7 ( L_{bulge}/5 \times 10^9 \Lsun)^{1.2}$. 

Because quasars were populous in the youthful Universe, but 
have mostly died out, the Universe should be populated with 
relic black holes whose 
average mass density $\rho_u$ matches or exceeds the mass--equivalent of
the energy density $u$ emitted by them \cite{chokshi}. 
The integrated comoving mass--equivalent density in quasar light (as emitted)
is 
\begin{equation}
\rho_u = {1 / (\epsilon c^2)}
 \int_0^\infty \int_0^\infty \, L \Phi_Q(L|z) dL \, {dt \over dz} \, dz
=            2 \times 10^5  
            \left({ 0.1 \over \epsilon} \right) \Msun \, \mpc^{-3}.  
\end{equation}
where $\Phi_Q$ is the comoving density of quasars of luminosity $L$, 
and $t$ is cosmic time and $\epsilon$ is  the radiative efficiency.  
This density can be compared to the luminous density in galaxies.  
Using Loveday's estimates of the parameters of a Schechter luminosity 
function 
\begin{equation}
\Phi_G(L)dL 
= \phi^* ({L \over L^*})^{-1} \, e^{-L/L^*} \, d({L \over L^*})
\end{equation}
with $\phi^* = 1.4 \times 10^{-2} h^3 \mpc^{-3}$
gives a luminous density of 
$j = 1.1 \times 10^8 \, \Lsun \, \mpc^{-3}$ 
\cite{loveday}, we obtain the ratio of the mass  
in relic MBHs to the light of galaxies ($h=0.8$):
\begin{equation}
  \Upsilon = 
     {\rho_u \over j} = 
      1.8 \times 10^{-3} \; \left( {0.1 \over \epsilon} \right)
         \left( {\Msun \over \Lsun} \right) . 
\end{equation}
We can compare the estimate of 
\cite{demogpaper} to the prediction of the total luminosity 
in quasars (above) by apportioning the quasar--predicted mass 
according to the mass of each galaxy.  The quasar light underpredicts 
the observed black hole masses by about a factor of 5, suggesting 
that a large fraction of black hole growth may occur at radiative 
efficiencies significantly less than $0.1$.  

\section*{An Attempt to Quantify the Merger Rate}

The previous section suggests that every galaxy hosts a massive 
black hole.  In the hierarchical model of galaxy formation elliptical 
galaxies form and grow as a result of generations of mergers of 
comparably massive progenitors.  The exact nature of the progenitors 
and the epoch of the mergers are both uncertain, but several lines 
of argument suggest that there is a high merger rate of galaxies 
containing bh's in our past light cone.  

The {\it number} density of galaxies above a luminosity of 
$0.01 \, L^*$ is (from eqn 2)
\begin{equation}
   	n_{0.01} = \phi^* \int_{0.01}^\infty x e^{-x} \, dx 
 = 5.6 \times 10^{-2} h^3 \mpc^{-3}
\end{equation}
  Multiplying this by the Hubble volume $4 \pi c^3/(3 H_0^3)$
gives an estimate of $6 \times 10^9$ galaxies 
in our past lightcone.  Dividing by $t_0 = 8 \times  10^{9} \yrs$ gives 
a merger rate since redshift $z = 1$ of $0.7 h \yrs ^{-1}$,   
if each galaxy undergoes one merger in that time.  
One might expect a comparable 
contribution to the merger rate from higher redshifts, at least 
up to $z \sim 3$ where the quasars are most numerous, suggesting 
that the massive bh population formed then or earlier \cite{nature}.  
There are several lines of argument that this merger rate is
reasonable.  The simplest approach to a galaxy merger rate is to 
use the Press--Schechter \cite{prs} formalism to estimate the change in the 
number of objects at a mass of about $10^{12} \Msun$ since $z = 3$.  
For a fairly standard normalization of $\sigma_8 = 1$ at present 
and a bias near 1 \cite{strauss-willick}, the number of collapsed 
objects has increased by about a factor of order unity in an 
$\Omega_0 = 1$ cosmology, and about 1/3 in a $\Omega_0 = .2$
Universe.  The merger rate at higher redshifts is higher.  
A better calculation based on semi--analytic galaxy formation 
models and ``conditional'' Press-Schechter formalism 
suggests a growth factor of $\sim 10$ \cite{somerville} since 
$z = 3$.  

A second argument can be made from the observations of the 
``Lyman break objects'', which suggests that the brightest objects 
seen at $z \sim 3$ are a factor of 10 less massive than bright 
objects today and considerably more numerous 
\cite{steidel,lowenthal,trager}.  

If the galaxies merge, do the massive black holes contained in them
merge as well?  Many of the calculations needed to answer this
question have been carried out in a somewhat different context
\cite{xu}.  Dynamical friction will carry a massive black hole of
$10^7 \Msun$ or more into the center of a host galaxy in less than a
Hubble time from far out in the galaxy.  Smaller black holes, cloaked
in sufficient numbers of bound stars from their parent pre--merger
galaxy, will similarly be carried to the center.  Two massive black holes
in the center of such a galaxy will form a hard binary which decays
increasingly slowly due to stellar scattering, until gravitational
radiation becomes important.  For galaxies with central densities like
the Milky Way, black holes more massive than $10^6 \Msun$ will reach a
high enough binding energy to decay by gravitational radiation in less
than a Hubble time.  A similar look at this problem in a variety of
galaxy types would be valuable.

Finally, there is some observational information that can be brought
to bear on the question of mergers of black holes in our past light
cone.  Although the observed mass density of supermassive black holes is
only 5 times greater than that predicted from the integral of the
quasar light, the {\it number} of black holes of $10^8 \Msun$,
corresponding to Eddington luminosities of $\sim 10^{46} \, {\rm
ergs/sec}$, is about $10^{-3} \mpc ^{-3}$ \cite{nature}, while the
number of quasars with luminosities of $10^{46} \, {\rm ergs/sec}$, at
the peak of quasar numbers at $z \sim  3$ is only about $10^{-6}
\mpc ^{-3}$ \cite{ssg}.  This discrepancy can be resolved in one of
two ways.  The obvious one is that quasars shine only for about $10^6
\yrs$. This seems implausible as they could then only accrete (even at
super Eddington rates) a few percent of their mass in this time, and
must gain the rest in a manner invisible to us.  Alternatively, they
may have merged a few times since the quasar era.  Even two
generations of merging (producing a factor of 4 change in mass of a
typical black hole since the quasar epoch) goes a long way to
resolving the ``numbers crisis'' because we must then identify the
quasars that powered the bright quasars at early epochs with much more
massive black holes today.  Since the galaxy luminosity function (and
by hypothesis, the bh mass function) falls exponentially at high
mass this modest growth factor serves to bring the numbers at high and
low redshift into line (see \cite{nature}).  

Thus it seems likely on several grounds that the supermassive black
hole population has undergone a few mergers since the quasar epoch.
If this is so, the bh merger rate in our lightcone could easily 
exceed $1/\yrs$.  
As noted in other talks at this meeting, these mergers should be
observable for masses of at least $10^6 \Msun$.  Since our best current
understanding is of yet higher mass black holes, it would be desirable
to maintain or improve LISA's performance at the lowest frequency,
where the heaviest objects will radiate.  On the other hand, LISA may
give us the best handle on the mergers of the low mass objects, and
may provide the only information we will get on the mergers of
protogalaxies before $z \sim 3$.  

I'm grateful to the ``Nukers'' ({E. A. Ajhar, R. Bender,
G. Bower, A. Dressler, S. M. Faber, A. V. Filippenko, K. Gebhardt,
R. Green, L. C. Ho, J. Kormendy, T. Lauer, J. Magorrian \&
S. Tremaine}) and also to John Bahcall, Pawan Kumar, J. P., Ostriker, 
Martin Rees and David Spergel, for discussions of these topics.  
I thank the Ambrose Monell Foundation, the Guggenheim Foundation and 
NASA for financial support.

\end{document}